\documentstyle[11pt,newpasp,epsf]{article}

\begin{document}

\title{Galaxy Interactions and Starbursts at High Redshift}

\author{R.S. Somerville$^1$, G. Rosenfeld$^1$, T.S. Kolatt$^2$, 
A. Dekel$^1$, J.C. Mihos$^3$ and J.R. Primack$^2$}
\affil{$^1$Racah Institute of Physics, Hebrew University, Jerusalem \\
$^2$ Physics Department, University of California, Santa Cruz \\
$^3$ Department of Astronomy, Case Western Reserve University}

\begin{abstract}
Using high resolution N-body simulations with hydrodynamics and star
formation, we investigate interactions and the resulting starbursts in
galaxies with properties typical of $z\sim 3$. We apply spectral
population models to produce mock-HST images, and discuss the observed
magnitude, color, and morphological appearance of our simulated
galaxies in both the rest-UV and rest-visual bands.
\end{abstract}


\section{Introduction}
An interest in galaxy interactions at high redshift can be motivated
from several directions. It has been shown that many of the observed
properties of galaxies at $z\sim 3$--4 (e.g., their number density,
luminosity function, colours, sizes, velocity dispersions, etc.) are
well-reproduced by a model in which these galaxies are predominantly
starbursts triggered by galaxy interactions (Somerville, Primack, \&
Faber 1999; SPF). In addition, the collision rate of dark matter halos
in high-resolution cosmological N-body simulations shows a marked
increase at earlier epochs, and the clustering properties of these
colliding halos at $z\sim3$ are similar to those of the observed
Lyman-break galaxies (Kolatt et al. 1999).

Previous numerical investigations of starbursts in interacting
galaxies (e.g. Mihos \& Hernquist 1994a; 1996) have assumed initial
properties typical of local galaxies. Many of these properties (e.g.,
gas content, surface density, disk-to-halo ratio) may be quite
different at high redshift. Using the same code as Mihos \& Hernquist,
we are carrying out an ongoing program of simulations to more fully
explore the parameter space with an emphasis on high-redshift
galaxies. A more complete version of these preliminary results will be
presented in Rosenfeld et al. (in prep.).

\section{Simulations}
We use the TREESPH code, including star formation using a Schmidt law
($\dot{\rho_{*}} = C \rho^N_{gas}$) as described in Mihos \& Hernquist
(1994b). We set the proportionality constant in this equation by
requiring the isolated galaxies to lie on the relation given by
Kennicutt (1998). Masses and mass ratios of typical colliding dark
matter halos at $z=3$ were determined using the simulations discussed
in Kolatt et al. (1999). For all calculations requiring the assumption
of a cosmology, we assume the same cosmological model as those
simulations, namely $\Omega_0=0.3$, $\Omega_{\Lambda}=0.7$, and $H_0 =
70$ km/s/Mpc. The mass and exponential scale radius of the stars and
gas in the galactic disk inhabiting a halo of a given mass at the
desired redshift are estimated using the semi-analytic models of SPF,
and are consistent with known properties of Lyman-break galaxies at
$z\sim 3$. The disks are assumed to be stable before the start of the
interaction. The results presented here are for an interaction between
equal mass, bulgeless galaxies, with halo mass $7.1\times 10^{11}
M_{\sun}$, stellar mass $7.2 \times 10^{9} M_{\sun}$, stellar scale
radius 1.7 kpc, gas fraction $0.5$ and gas scale radius 3.4 kpc. All
other properties (relative inclination, orbit, etc.), are the same as
the fiducial case of Mihos \& Hernquist (1996; MH96).

\section{Results}
\begin{figure}
\plottwo{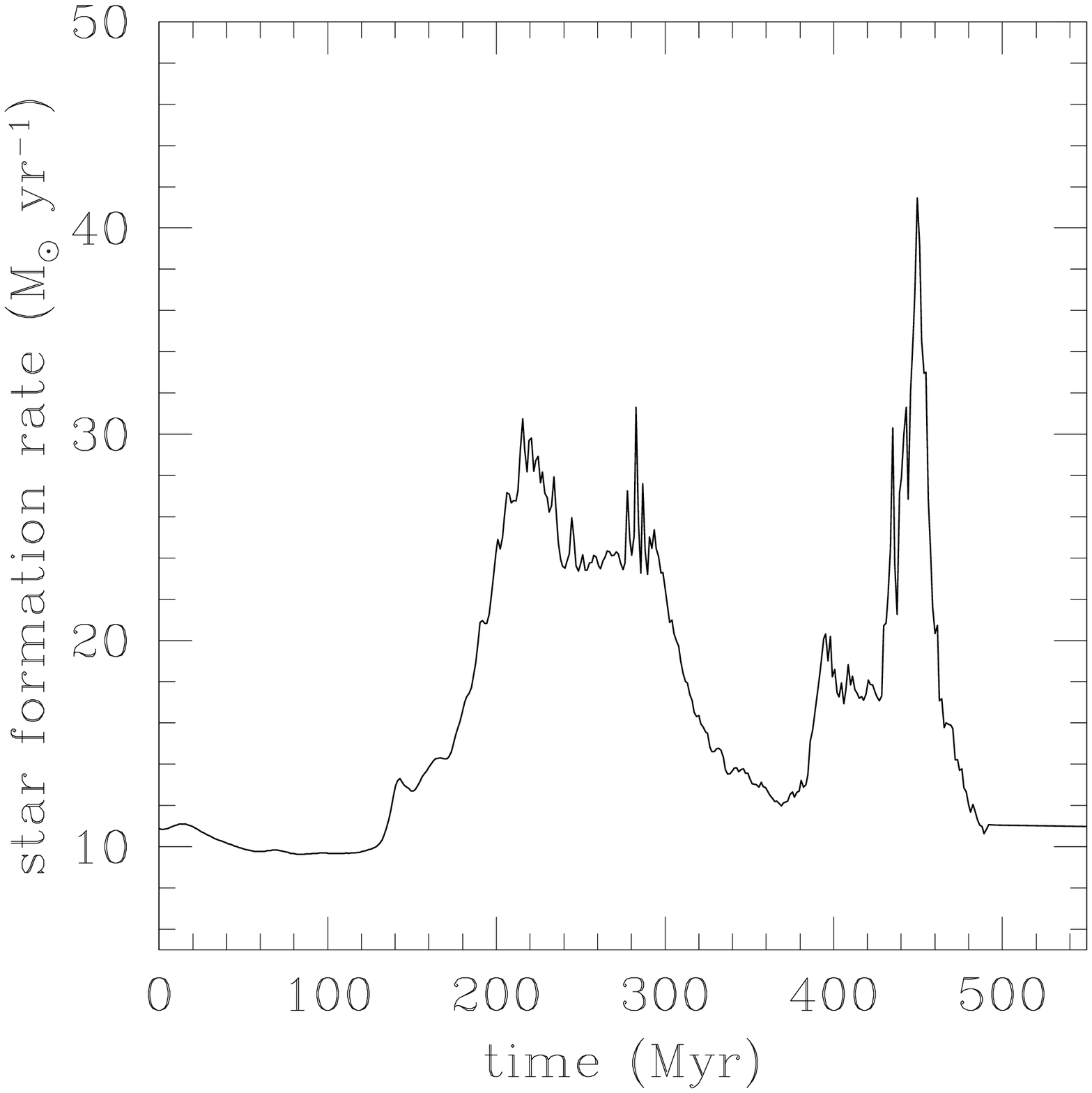}{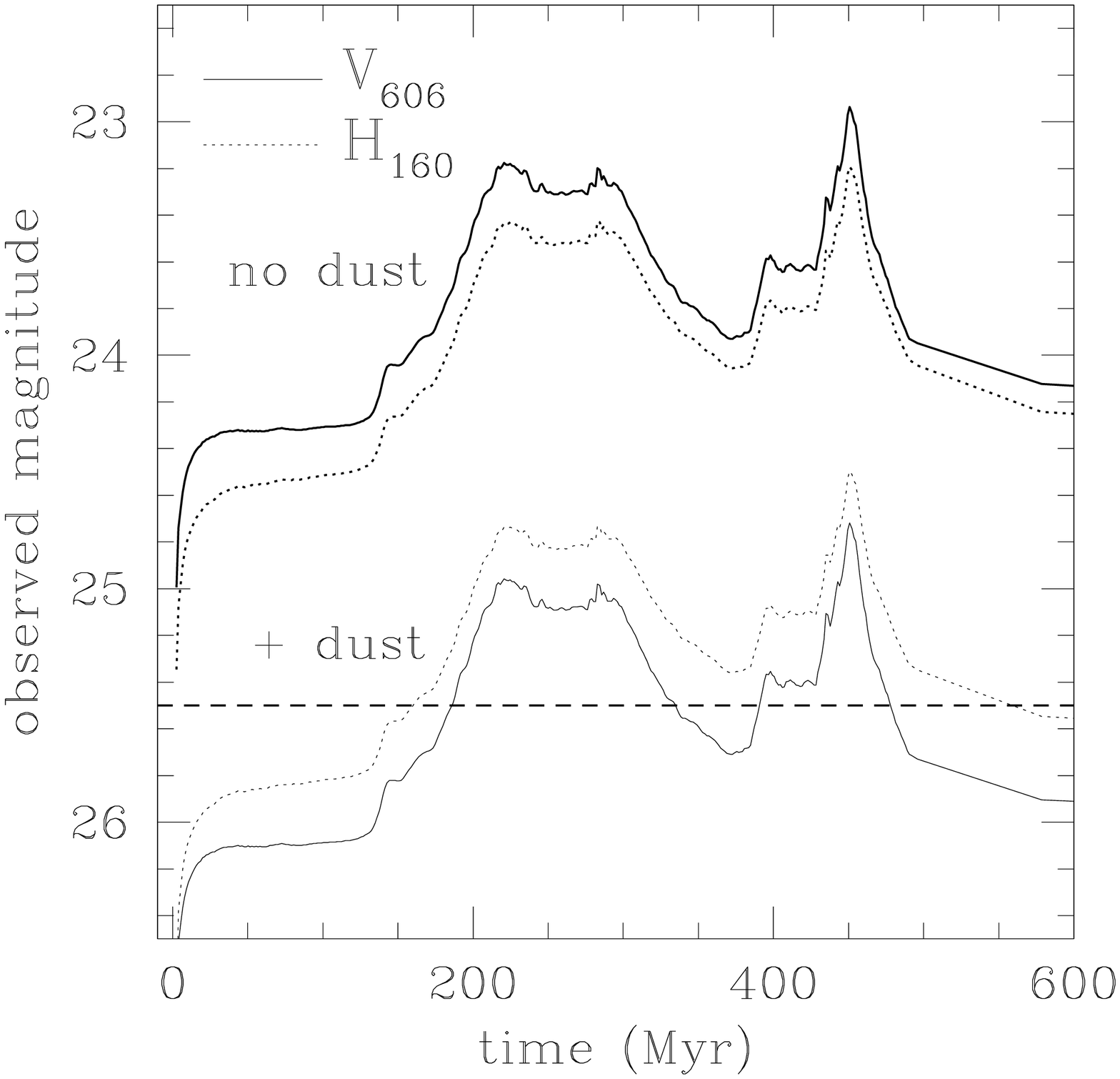}
\caption{Star formation rate (left) and observed magnitude at $z=3$ as 
a function of time since the start of the simulation. In the right
panel, the bottom curves include an estimate of the effects of dust
(see text). } \label{fig-1}
\end{figure}

The star formation rate over the course of the interaction is shown in
figure 1 (left panel). The behaviour is qualitatively similar to the
results of MH96. We convolve this star formation history with stellar
population models (GISSEL98, Bruzual \& Charlot in prep.) to obtain
the apparent magnitude of the galaxy at $z\sim3$ (figure 1, right
panel). We have used the solar metallicity models with a Salpeter
IMF. We show the magnitudes in the WF/PC F606W ($V_{606}$) and NICMOS
F160W ($H_{160}$) filters (AB system), which probe the rest-frame
$\sim 1500
\AA$ and $4000 \AA$ part of the spectrum at $z=3$. The top set of
curves in the right panel neglects dust extinction; the bottom set
shows the result of including a correction of a factor of $\sim 5$ at
1500 \AA, as suggested by recent observational estimates of typical
extinction corrections in bright LBGs (Meurer et al. 1999; Steidel et
al. 1999), and assuming a Calzetti attenuation curve (Calzetti et
al. 1996). With this level of dust extinction, the bursting galaxy
would be visible at present spectroscopic limits for about 200 Myr,
whereas in the absence of the burst the galaxy would have been well
below the detection limit. Note that in the absence of dust reddening,
the $V_{606}-H_{160}$ color of the galaxy during the burst is quite
blue (-0.5 -- 0), but after the dust correction, typical values (0.5
to 1.0) are reasonably consistent with observations.

Figure 2 shows how the interacting galaxy would appear if observed at
$z\sim3$ by HST (in the absence of noise or sky background) at various
times during the merger. Left panels show the WF/PC $V_{606}$ filter
with a pixel size of 0.04 arcsec; the right panels show the NICMOS
$H_{160}$ filter with 0.08 arcsec pixels. From top to bottom, the
galaxies are shown at the beginning of the simulation, during the
first interaction of the galaxies, after the galaxies have separated
again, and finally after the final merger. These images can be matched
up with the star formation and total magnitude curves shown in
figure~1 by multiplying the simulation time units shown on the figure
by a factor of 8.3 to convert to Myr.

If in fact the observed galaxy population becomes increasingly
dominated by merging starburst systems at higher redshifts, one would
expect to observe larger fractions of galaxies with disturbed
morphologies and significant substructure. We plan to use synthesized
images such as figure~2 to develop statistics to quantify these
effects, and eventually to test whether the morphological properties
of observed high-redshift galaxies are consistent with the collisional
starburst scenario.

\begin{figure}
\includegraphics{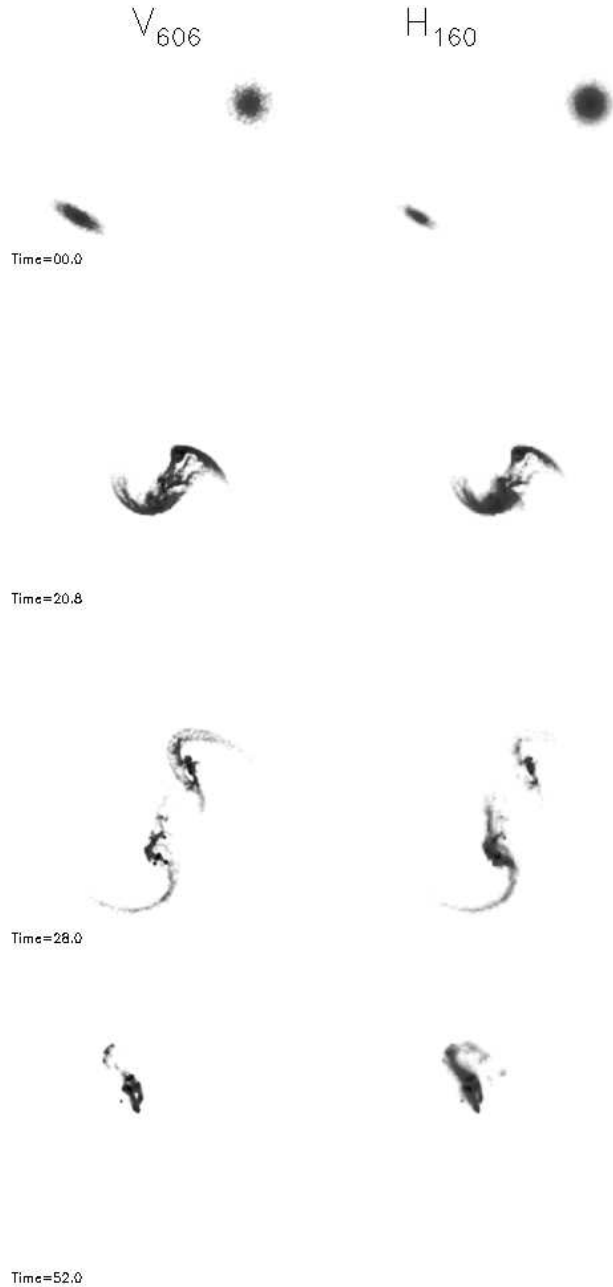}
\vspace{18truecm}
\caption{Mock HST images of simulated starburst galaxies at
$z\sim3$ at different times during the merger. Left panels shows the
$V_{606}$ filter (rest $\sim1500 \AA$) and right panels show the
NICMOS $H_{160}$ filter (rest $\sim4000 \AA$). The lightest grey
corresponds to a surface brightness of 28 and the darkest to 18
magnitudes arcsec$^{-2}$. } \label{fig-2}
\end{figure}

\end{document}